# Reconfigurable optical computing through scattering media with wavefront shaping

*Zhipeng Yu, Yuchen Song, Tianting Zhong, Huanhao Li, Wei Zheng, and Puxiang Lai[*]*

Dr. Z.Yu, Y.Song, T.Zhong, Dr. H.Li, Dr. P.Lai
Department of Biomedical Engineering, Hong Kong Polytechnic University, Hong Kong SAR
Shenzhen Research Institute, Hong Kong Polytechnic University, Shenzhen 518057, China
E-mail: puxiang.lai@polyu.edu.hk

Prof. W.Zheng
Research Laboratory for Biomedical Optics and Molecular Imaging, Shenzhen Institutes of Advanced Technology, Chinese Academy of Sciences, Shenzhen 518055, China

**Abstract:** Optical logic gates are fundamental blocks of optical computing to accelerate information processing. While significant progress has been achieved in recent years, existing implementations typically rely on dedicated structures that are predesigned to modulate the phases and intensities of input optical beams accurately for specific logic functions. Thus, these optical gates usually lack reconfigurability and are incapable within or through optically opaque media/environment, such as heavy fog and turbid water. In this work, we propose reconfigurable optical logic operations through scattering media with transmission matrix-based wavefront shaping. A light beam is reflected by a spatial light modulator that is divided into several subregions functioning as logic units, with each displayed with predetermined wavefronts via transmission matrix-based wavefront shaping. Each modulated wavefront transmits through the scattering medium to form a desired light field. The interference of these light fields generates bright optical focus at pre-assigned locations in space, representing different logic states. As a proof of concept, we experimentally demonstrate five basic logic functions (AND, OR, NOT, NAND, NOR). As the transmission matrix of the scattering medium/system can be measured instantly to adapt to environment perturbation, the method opens new venues towards reconfigurable optical computing in a dynamically complex environment.

## 1. Introduction

Optical computing, using photons instead of electrons, has attracted significant research interests in recent years because photons can accomplish the same goals as digital electronics, but in a more efficient or advantageous manner.[1-6] The benefits, such as large bandwidth, ultrahigh speed, low energy consumption, low heat generation, and low crosstalk, make it potential in many scenarios, particularly those involving high-throughput and on-the-fly data processing such as artificial intelligence.[7-9] Optical logic gates are the fundamental building blocks of optical computing; its exploration is thus particularly important. Existing optical logic components are mostly based on waveguide[10-16] or metasurfaces. [17-18] In the first category, logical states are achieved mainly through constructive and destructive interference among input light beams, and precise control of the basic properties (phase, polarization, and intensity) of individual optical signals transmitting into the waveguide should be observed. In the second category, in order to function as optical logic gates, a plane wave is usually used as the input signal to be delivered through a preset group of subregions of complex structures.

In both categories, the implementations typically rely on dedicated structures that are predesigned and fabricated to modulate the phases and intensities of input optical beams accurately for specific logic functions. Thus, existing optical gates usually lack reconfigurability[14, 19-20] and can only perform with ballistic or quasi-ballistic light. It becomes challenging or has not yet been proved to be compatible with diffused light, which is indispensable for information transmission or object identification in some complex environments, such as automatic driving in heavy fog or rain[21] and optical communication in turbid water.[22-23] Optical computing within or through scattering media is desired in these scenes yet remains unexplored.

In this work, we propose to tackle the challenge with the assist of wavefront shaping, a promising tool conceived by Vellekoop and Mosk[24] to manipulate scattered light beyond the optical diffusion limit.[25-30] As known, when light transmits within or through scattering media, such as a ground glass or biological tissue, the optical wavefront is scrambled due to scattering, forming visually random speckle patterns. This seemingly random process is actually deterministic in a specific time window,[31] within which a transmission matrix (TM) model can be used to bridge the input and output optical fields.[30] For a strong scattering medium, the distribution inhomogeneity of refractive index is on a sub-wavelength scale, which supports a large number of optical modes and hence a TM of large dimension. Moreover, the random distribution of scatterers within the medium induces asymmetry, resulting in high rank to the TM. If the incident light is modulated by a spatial light modulator (SLM) with a wavefront that is inverse to the TM or other well-chosen wavefronts, an optical scattering medium can be used to perform a wide variety of functions that are otherwise impossible, such as optical focusing,[32] image transmission,[30] programmable quantum network[3], and reconfigurable linear operation[33] by using diffused light.

Inspired by these exciting features of TM-based wavefront shaping, in this study we employ a digital micromirror device (DMD, a type of SLM) to be loaded with precalculated wavefront pattern to modulate light before it transmits through a ground glass to form logical states. The DMD aperture is divided into several subregions, as illustrated in **Figure 1**, that correspond to binary input digits ("0" and "1"), logic types (AND, OR, NOT, NAND, NOR), and a common reference region, respectively. The sub-TM of the scattering medium corresponding to each subregion is determined by the TM-based wavefront shaping approach.[34-35] Two randomly selected small regions on the output plane (*e.g.*, the camera sensor) serve as the output optical logic states "0" and "1"; light projected from each DMD subregion through the scattering medium can be refocused to these two regions, individually or simultaneously, with two different phase states ("0" indicates in phase and "π" indicates in opposite phase) with respect to the speckle pattern projected from the common reference region on the DMD. Built upon constructive or destructive interference among the focuses, different combinations of subregions on the DMD loaded with precalculated wavefronts can perform different logic functions. An inverse design adopted in existing optical logic gates is usually incompatible with reconfigurability. But our method can bypass this drawback: when the system is considerably perturbed or changed, which causes large variations to the TM of the scattering medium/system, another round of rapid TM-based wavefront shaping can be performed, recovering the states of the optical logic gates. In addition to optical computing, the demonstrated capability also shows potentials in other fields like optical encryption[36] and optical micro manipulation.[37-38]

## 2. Method

Figure 1 illustrates the schematic of the proposed optical logic gates through scattering media. The input layer of the logic gates is a DMD, which is used in this study with two aims. Firstly, it serves as a spatial light modulator that provides binary amplitude modulation to the wavefront of the input light. Secondly, it functions as a fast optical (spatial) switch as the direction of the reflected light from the DMD is subject to change when the DMD pixels are at different states ("on" or "off"). As shown in the inset of Figure 1, the DMD screen is divided into two parts: the central region consisting of nine working subregions that control the binary digit or the logic type of the input, and the peripheral region (in blue) serving as the common reference region. Among the nine working subregions, four are for input binary digits, marked as digit "0" or "1" (subregions of the same column belong to one set), and the rest represent logic types (AND, OR, NOT, NAND, NOR). These nine working subregions are equal in area and the ratio of the pixel number between the reference region and each working subregion is about 8. On the output plane (*i.e.*, camera sensor), two regions are selected to represent logical states "0" and "1". For details of the optical setup used in this study, please refer to Supporting Information 1.

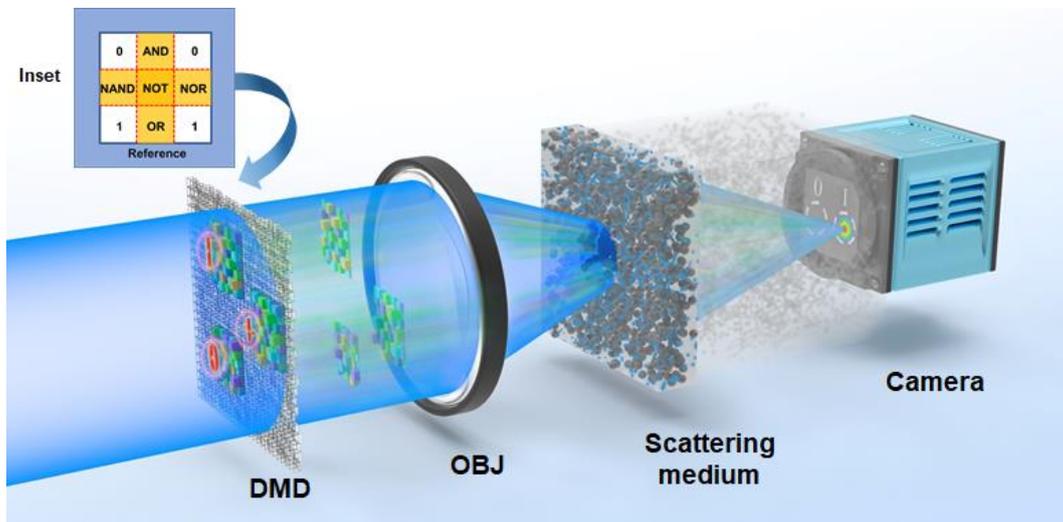

**Figure 1.** Schematic diagram of the proposed optical logic gates. The encoder is a digital micromirror device (DMD) of reflection type. For simplicity, it is shown here in transmission mode. OBJ: objective lens. Scattering medium is a ground glass diffuser in this manuscript. The output optical filed is recorded by a digital camera. The inset illustrates the arrangement of subregions on the DMD. Subregions marked in yellow represent the logic type control units, subregions marked in white represent the binary input digit units, and the subregion marked in blue serves as the common reference region. In operation, one of the control units and one or two of the input digit units are selected and loaded with predetermined phase masks, while all other subregions remain

inactive ("0" values for all of their pixels).

The proposed optical logic gate can be analog to electronic logic gates, which are achieved by using different circuit structures. For example, in electronics, the "AND" or "OR" gate can be achieved using a diode-based circuit, and the "NOT" gate can be achieved using a transistor-based circuit. Once electronic components in the circuit are installed for a specific operation, these components cannot be changed. Of course, the voltage level of input signals can be changed to achieve different operations. In this regard, it is very similar in our system. For example, the function of the subregion representing "AND" logic type loaded with the precalculated binary pattern is analog to the function of a diode in electronic "AND" logic gate. Similar explanations can be extended for other logic types. DMD subregions representing binary digit "0" and "1" are analog to the low and high voltage levels, respectively. To be specific, for any logic operation, only those subregions involved in the operation remain "open" and are loaded with the precalculated wavefronts, while other subregions are "closed" with the corresponding pixels switched to "off". The output images are recorded by a CMOS camera. Taking logic operation "0·1" as an example. As illustrated in Figure 1, three subregions marked as "0" (from the leftmost column), "AND", and "1" (from the rightmost column), respectively, are selected and displayed with precalculated wavefronts; these three subregions are set to be "open" while other subregions are "closed". The output logical state is "0", so on the output plane a bright focus is formed at the region corresponding to logic state "0". Other optical logic operations have similar procedures, except for logic operation "NOT". The NOT logic gate is an inverter of a digital signal, converting a "0" to a "1", and vice versa. For optical logic operations "$\bar{0}$" and "$\bar{1}$", only two DMD subregions are "open": one corresponds to the input binary digit and the other represents logic type "NOT".

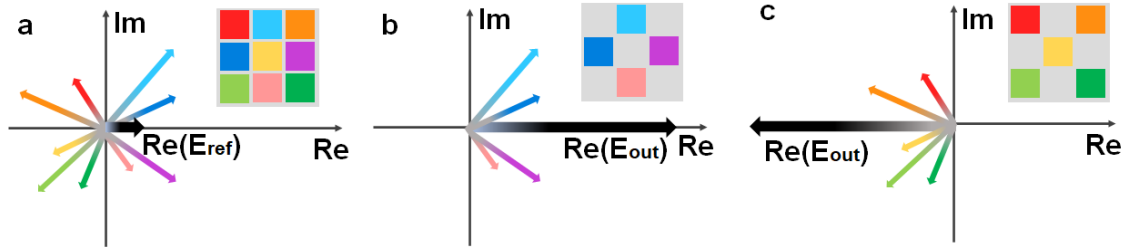

**Figure 2.** (a) Optical transmission through a scattering medium without binary modulation. The optical field $E_{Ref}$ at the target is the sum of electric fields coming from all optical channels, and it is used as the reference. (b) After measuring the binary transmission matrix (TM) of the medium, an enhanced focus at the target, which is in phase with the reference, can be achieved by blocking the channels that interfere destructively with $E_{Ref}$. (c) An enhanced focus at the target, which is in opposite phase with the reference, is achieved by blocking the channels which have constructive interference with $E_{Ref}$.

The method is achieved based on constructive or destructive interference among focuses formed by different subregions, and hence the key is how to modulate the phase and intensity of each individual optical focus accurately. In Ref. [35], the speckle field from the scattering medium when all DMD pixels (channels) are switched "on" is used as the reference during the TM measurement process. For a fully scattering medium, the phase profiles of the resultant optical field are randomly distributed, suggesting that the phase and intensity of optical field resulting from different input channels are randomly distributed, as illustrated in **Figure 2**a. To achieve constructive interference at the output plane, which is needed to obtain an optical focus, we need to block channels that lead to destructive interference with the reference $E_{Ref}$. That is to open the channels with output phase in the range of ($\varphi_{ref} - \pi/2, \varphi_{ref} + \pi/2$) while block the other channels, where $\varphi_{ref}$ is the phase of the reference. [35] As a result, the optical focus is in phase with the reference speckle field, as shown in Figure 2b. On the contrary, if we block channels that have constructive interference with the reference, a bright focus with an opposite phase to the reference speckle field can be obtained, as shown in Figure 2c. We can get optical focuses from different subregions in the same region on the output plane, but there is no explicit phase difference among their phases as their reference speckle fields are different. As the logic states are achieved based on constructive or destructive interference among different focuses generated by different working subregions at the same position on the output plane, furtherly bridging the phase relationship between different focuses is crucial. Therefore, we modify the aforementioned method by introducing a common reference subregion on the DMD (the blue region in Figure 1). There are a lot more pixels in the reference region than individual subregion to ensure that the reference contribution dominates in the interfered speckled field. All pixels in the reference subregion are switched "on" to project light into the scattering medium during the entire TM measurement process for all working subregions (control units). These pixels are switched "off" when the TM measurement ends. We use a Hadamard matrix of a size of $2^{10} \times 2^{10}$ to calculate the sub-TM for each subregion (128×128 pixels) with 4×4 DMD pixels merged as one mega pixel. Details about the TM calculation are provided as below.

We use a transmission matrix method to relate the relationship of electric fields between the input and output channels. Let us denote $T = \begin{bmatrix} t_{1,1} & \cdots & t_{1,n} & \cdots & t_{1,N} \\ \vdots & & \vdots & & \vdots \\ t_{k,1} & \cdots & t_{k,n} & \cdots & t_{k,N} \\ \vdots & & \vdots & & \vdots \\ t_{K,1} & \cdots & t_{K,n} & \cdots & t_{K,N} \end{bmatrix}$ to be the transmission matrix (TM) of the scattering medium, and $T_k = [t_{k,1} \cdots t_{k,n} \cdots t_{k,N}]$ to be the $k^{th}$ row vector of T, connecting the $k^{th}$ input and output channels. Here, $N$ is the number of control units on the DMD, and K is number of output channels. Let us further denote $E_{out} = [e_{out\_1} \cdots e_{out\_k} \cdots e_{out\_K}]$ as the electric field of focuses at all output channels, and $e_{out\_k}$ as the electric field of the focus at a desired output channel ($k^{th}$). $e_{out\_k}$ is the sum of contribution from all input channels and is given by

$$e_{out\_k} = T_k (D_k e_0)^\dagger, \tag{1}$$

where $e_0$ represents a plane wave which will illuminate the DMD.

$$D = \begin{bmatrix} d_{1,1} & \cdots & d_{1,n} & \cdots & d_{1,N} \\ \vdots & & \vdots & & \vdots \\ d_{k,1} & \cdots & d_{k,n} & \cdots & d_{k,N} \\ \vdots & & \vdots & & \vdots \\ d_{K,1} & \cdots & d_{K,n} & \cdots & d_{K,N} \end{bmatrix}$$ is the binary modulation matrix, and symbol " † " represents the transpose of the matrix, $D_k = [d_{k,1} \cdots d_{k,n} \cdots d_{k,N}]$ is the $k^{th}$ row vector of D and it is the pattern to be displayed on the DMD to produce a focus at the $k^{th}$ output channel. Then we can calculate the intensity of focus at the designated output channel:

$$I_{out\_k} = |e_{out\_k}|^2 = \left| \sum_{n=1}^{N} t_{k,n} d_{k,n} e_0 \right|^2, \tag{2}$$

where $t_{k,n}$ follows a circular Gaussian distribution. [24] To maximize the intensity at the targeted output position (in phase with the common reference), as shown in Figure 2, control units on the DMD can be modulated to obtain constructive interferences with the reference at targeted positions by switching on constructive channels and blocking others. $d_{k,n}$ can be determined based on the value of $t_{k,n}$, following a criteria introduced in Ref. [40]:

$$d_{k,n} = \begin{cases} 1 & \text{Re}(t_{k,n} e_0) \geq 0 \\ 0 & \text{Re}(t_{k,n} e_0) < 0 \end{cases}. \tag{3}$$

Also to be noted that in our method, control units on the DMD are also modulated to form destructive interferences with the common reference field at targeted positions by switching on the destructive channels and blocking others to maximize the intensity at the targeted output position (in opposite phase with the reference). These have been described in detail in the "Results" session. The main goal of doing so is to determine the sign of every element of $\text{Re}(T)$. If we use Hadamard basis to represent the input, the output light field for different Hadamard modes on the DMD is

$$[E_{H1} ... E_{Hn} \cdots E_{HN}] = T[h_1 \cdots h_n \cdots h_N]^\dagger e_0, \tag{4}$$

where $E_{Hn}$ is the output field for $n^{th}$ Hadamard mode, which is a $K \times 1$ vector. $h_n$ is the $n^{th}$ Hadamard mode, which is a $N \times 1$ vector. As a Hadamard matrix has the inverse property of $\mathbf{HH}^T = \mathbf{NI}$, $\text{Re}(Te_0)$ can be calculated by

$$\text{Re}(Te_0) = \frac{1}{N} \left[ \text{Re}([E_{H1} \cdots E_{Hn} \cdots E_{HN}]) \right] [h_1 \cdots h_n \cdots h_N]^\dagger. \tag{5}$$

It is challenging to directly measure $\text{Re}(E_{Hn})$ as elements of the Hadamard matrix are either -1 or 1. To solve the issue, we use a same method as Ref. [34] by adding a uniform reference input. Two binary DMD patterns ($r_n^\pm$) which only contain 0 and 1 are generated for each $h_n$ added with a uniform reference

$$r_n^\pm = \frac{1}{2}(h_1 \pm h_n). \tag{6}$$

The uniform reference input field is the first Hadamard basis vector, $h_1 = [1 \cdots 1]$. Note that in our method, an extra reference region is introduced, as shown in Figure 1, to bridge the phase relationship among focuses generated from different subregions. The intensity of the resultant output light field on the output plane can be expressed by

$$I_{total\_n}^\pm = \left|Tr_n^\pm e_0 + E_{ext}\right|^2 = \left|E_{ref} \pm \frac{1}{2}Th_n e_0\right|^2 = \left|E_{ref}\right|^2 + \frac{1}{4}\left|E_{Hn}\right|^2 \pm \text{Re}(E_{ref}^* E_{Hn}), \tag{7}$$

where $E_{ref} = \frac{1}{2}Th_1 e_0 + E_{ext}$ is the electric field of the resultant output light field from the uniform reference input field ($\frac{1}{2}Th_1 e_0$) and the extra reference region ($E_{ext}$). $E_{Hh}$ is the electric field for input of the $n^{th}$ Hadamard basis vector, which can be derived from Equation 7:

$$\text{Re}(E_{Hn}) = \frac{\frac{1}{2}(I_{total\_n}^+ - I_{total\_n}^-)}{\sqrt{I_{ref}}} \cong \frac{\frac{1}{2}(I_{total\_n}^+ - I_{total\_n}^-)}{\sqrt{I_{ext}}}, \tag{8}$$

where $I_{ref}$ is the optical intensity of the resultant output light field from the uniform reference input and the extra reference region, and $I_{ext}$ is the optical intensity of the output light field from the extra reference region. As the area of the extra reference region is 8 times larger than that of individual working region, $I_{ext}$ is much larger than that from the working region, thus we have $I_{ref} \cong I_{ext}$. At last, the condition in Equation 3 can be obtained from Equation 5 and 8. By introducing a large reference region, after wavefront shaping, the phase of focuses generated by each DMD subregions can be precisely controlled: it can be either in phase or in opposite phase with reference speckle field resultant from the reference region (Experimental results can be referred to Supporting Information 2).

Note that if only one focus is generated from one DMD subregion, the logic operations cannot be completed. For example, for operation "$\bar{0}$" with logic output of "1", there should be constructive interference between optical focuses formed by DMD subregions corresponding to control units "0" and "NOT" in the designated region on the output plane for logic state "1" but destructive interference in the designated area for logic state "0". Therefore, two focuses of specific intensity and phase in the designated areas on the output plane should be generated synchronously by one DMD subregion. For this reason, an intersection operation is adopted in this study to generate two focuses synchronously with controllable intensity and phase (in phase or in opposite phase with the common reference) in the designated regions on the output plane by one DMD subregion (detailed results can be referred to Supporting Information 3). The procedure of the intersection operation is described below: two binary patterns, represented by row vectors $X = [x_1 \cdots x_n \cdots x_N]$ and $Y = [y_1 \cdots y_n \cdots y_N]$ (N is the pixel number), are selected to be displayed on individual subregion to generate an optical focus respectively in designated areas on the output plane after the TM is determined. In order to generate two focuses synchronously, a new pattern $Z=[z_1 \cdots z_n \cdots z_N]$ can be calculated based on an intersection operation between these two DMD patterns, that is $z_n = \begin{cases} 1, & \text{when } x_n \times y_n = 1 \\ 0, & \text{when } x_n \times y_n = 0 \end{cases}$. The intersection operation is effective and reliable to control the phases and intensities of the focuses formed by individual DMD subregions, which is the foundation of achieving the proposed optical logic operations.

## 3. Results

### 3.1 Arrangement of focuses formed by individual DMD subregions

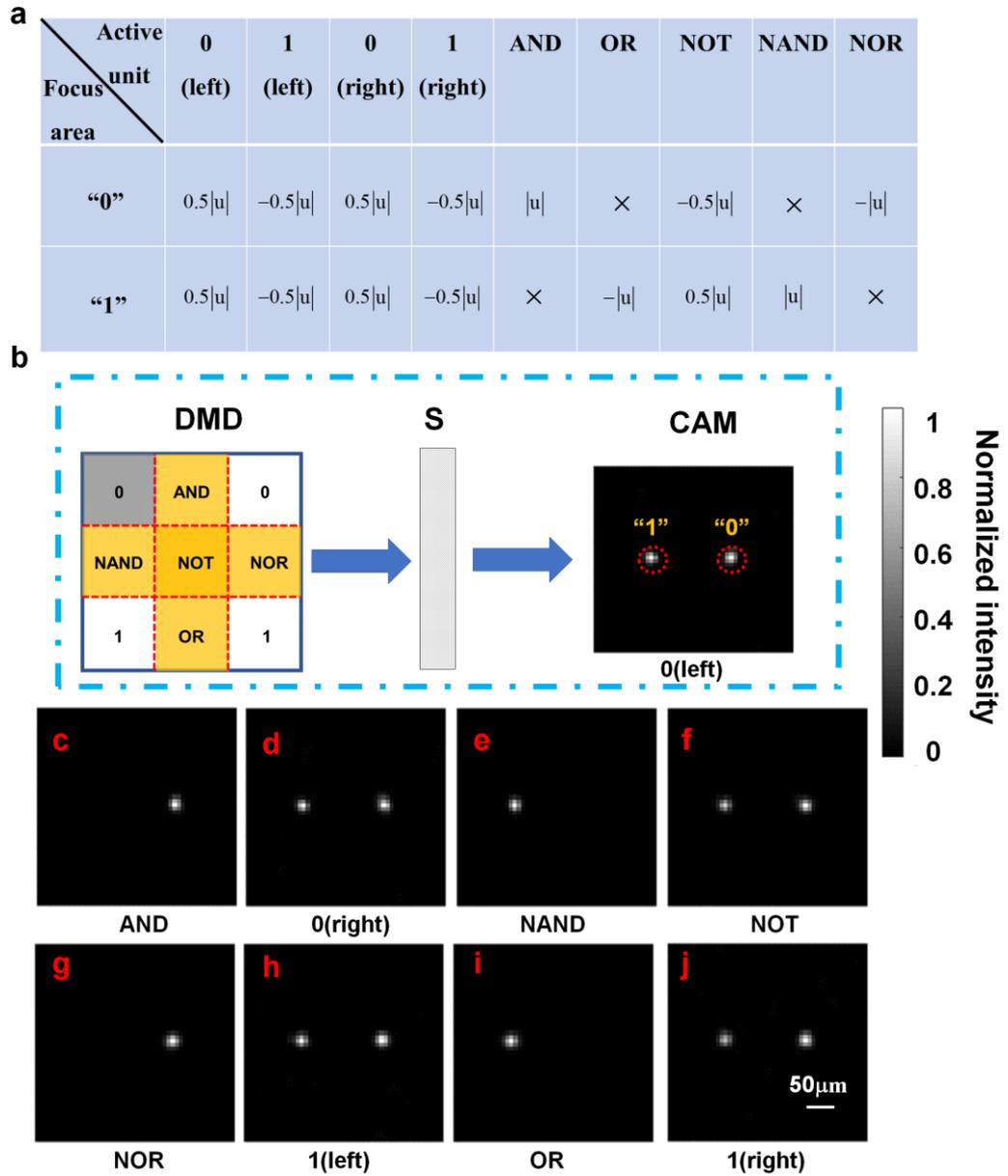

**Figure 3.** (a) Design arrangement of focuses formed by the nine DMD working subregions via numerical computation. Symbol "|u|" represents the absolute amplitude value of the electric field of single focus generated

from individual DMD subregions, and "0.5|u|" means there is an intersection operation on the corresponding subregions, leading to dual focuses, whose absolute amplitude value of the electric field is reduced by half. "×" indicates there is no focus formed at the designated area. "-" means that focus is in opposite phase with the common reference; without "-" means the focus is in phase with the reference. (b-j) Experimental verification of the desired focus arrangement. As an example, in (b), a binary digit control unit ("0" on the left column of the DMD, marked in grey), is selected. Two focuses are formed in the designated areas at the output plane behind the scattering medium (marked in yellow on the output plane). Blue arrows indicate the transmission of light. Results with other control units are shown in (c-j). Abbreviations: DMD, digital micromirror device; S, scattering medium (ground glass); CAM, camera. Intensities are normalized to the maximum intensity of each Figure.

An inverse design approach was adopted. First, we calculated the desired intensity and phase (0 or π) of each focus for a set of basic logic functions (AND, OR, NOT, NAND, NOR) (the calculation process can be referred to Supporting Information 4). The solution is given in **Figure 3**a and explained in detail in the caption. These results were used to calculate the wavefront of each DMD subregion in reverse. Note that two randomly picked areas on the output plane serve as logic states "0" and "1". The experimental verification of the design arraignment is shown in Figure 3b-j. As an example (Figure 3b), when only the subregion corresponding to "0" on the left column of the DMD (marked in grey) is switched "open" and loaded with the precalculated wavefront, two optical focuses can be observed in both selected areas. Figure 3c-j display the output images when other subregions are "on" (one active subregion at one time) and loaded with the corresponding precalculated wavefront. As seen, all experimental results agree well with the simulated sets in Figure 3a. The enhancement ratio defined as the ratio between the optimized intensity and the average intensity before optimization[24] for cases with one focus (Figure 3c, e, g, and i) is about 70 (43% of the theoretically predicted value of 164); for cases with two focuses (Figure 3d, f, h, and j), the enhancement ratio is about 30. The difference is reasonable as the number of effective channels contributing to the focuses are reduced by half due to the intersection operation, which has been discussed earlier.

### 3.2 Experimental demonstration of basic logic functions

Now, let us move onto the demonstration of the proposed optical logic gates. The DMD subregions will display the corresponding calculated wavefronts after the sub-TMs are measured for logic operations. In this process, the DMD serves as a wavefront modulator and a fast spatial switch (with a switching speed of up to 23 kHz). For a given logic operation, the selected subregions are set to be "on" while others are "off", in which the DMD is used as a spatial switch. Figure 4a-e show the DMD deployment for different optical logic gates. Inputs A and B are chosen between two subregions representing one set of binary digits ("0" or "1") in the leftmost column and rightmost column, respectively. For AND, OR, NAND, and NOR logic gates, Input A, Input B, and the subregion representing the corresponding logic type are activated, as shown in **Figure 4**a-e. For NOT logic gate, either Input A or Input B is activated besides the subregion representing NOT logic type, as shown in Figure 4c. Activated subregions will be loaded with the corresponding wavefronts that are precalculated based on the TM method discussed earlier. For a given logic operation, the logic output state ($Q_{AND}$- $Q_{NOR}$) can be either 0 or 1. Figure 4f shows the precalculated binary patterns on three subregions which correspond to binary digit control units "0" (left column) and "1" (right column) as well as the logic type control unit "AND", respectively, for the exampled "AND (0,1)" logic operation.

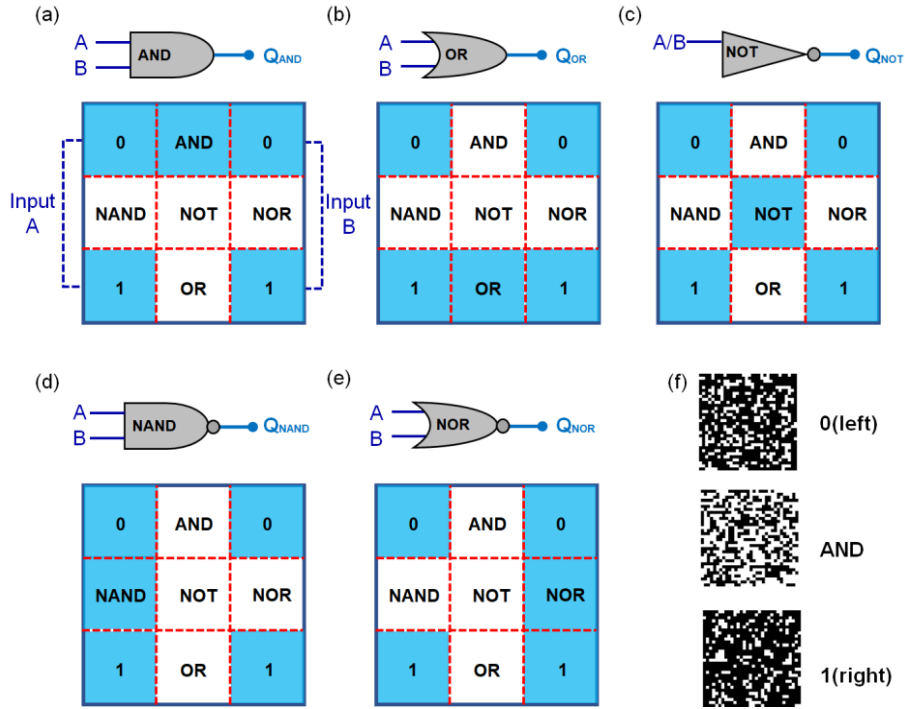

**Figure 4.** DMD subregion arrangement for AND (a), OR (b), NOT (c), NAND (d), and NOR (e) optical logic gates. Subregions marked in blue are in standby to be activated and loaded with the precalculated binary patterns in all logic gates. Input A is chosen from one set of binary digits ("0" or "1") on the leftmost column, and Input B is chosen from the other set of binary digits ("0" or "1") on the rightmost column. $Q_{AND}$-$Q_{NOT}$: logic output states. (f) The calculated binary wavefronts (pixels in white are switched "on") to be displayed on DMD Subregions "0" (Input A), "1" (Input B), and "AND", respectively, for optical logical operation AND (0,1). Each activated DMD subregion generates a corresponding optical field as shown in Figure 3.

20 sets of experimental results based on the five basic logic functions (AND, OR, NAND, NOR and NOT) and associated truth tables are shown in **Figure 5**. Switching on a desired group of subregions will result in appropriate optical logic output. For example, the AND gate registers $Q_{AND}$ = 0 for the exampled "AND (0, 1)" logic operation shown in Figure 4f. As defined, if a bright focus is formed at the region representing output logic state "1" on the output plane, the logic operation registers Q=1, and vice versa. As shown, every profile of red lines representing the experimental outputs for each mode behavior in Figure 5a-e agrees well with that of the respective truth table entry (Figure 5f) for ideal logic gates. According to the inverse design, the desired focus intensity contrast ratios of logic operations AND (0,0), OR (1,1), NAND (0,0), and NOR (1,1) are 6dB. For other logic operations, the contrast ratios shall be infinity as long as the focal peak to background ratios are sufficiently large to suppress the speckle background and most diffusive photons are focused onto designated regions. In our experiment, the intensity contrast ratios are 7.5 dB, 7.3 dB, 8.2 dB, and 8.2 dB, respectively, for logic operations AND (0,0), OR (1,1), NAND (0,0), and NOR (1,1). These are slightly better than the designed contrast ratios, which may attribute to two aspects. First, the focal peak enhancement ratio is not uniform on the whole output plane for light from one subregion [35] and the wavefront intersection affects the intensity of two focuses with different efficiencies. As a result, two focuses originating from the same subregion may have different focal intensities. Second, the focal peak enhancement ratios are also not the same at the same area on the output plane for different DMD subregions. As a consequence, other logic operations have decreased contrast ratios, which are merely more than 9.5 dB rather than infinity in our experiment. For example, for all "NOT" operations (Figure 5e), contrast ratios are larger than 13.9 dB. To further demonstrate the reliability of our method, 200 pairs of areas on the output plane were chosen in experiment as the focuses formed by light from individual DMD subregions and to represent the binary "1" and "0" outputs. Statistics can be referred to Supporting Information 5.

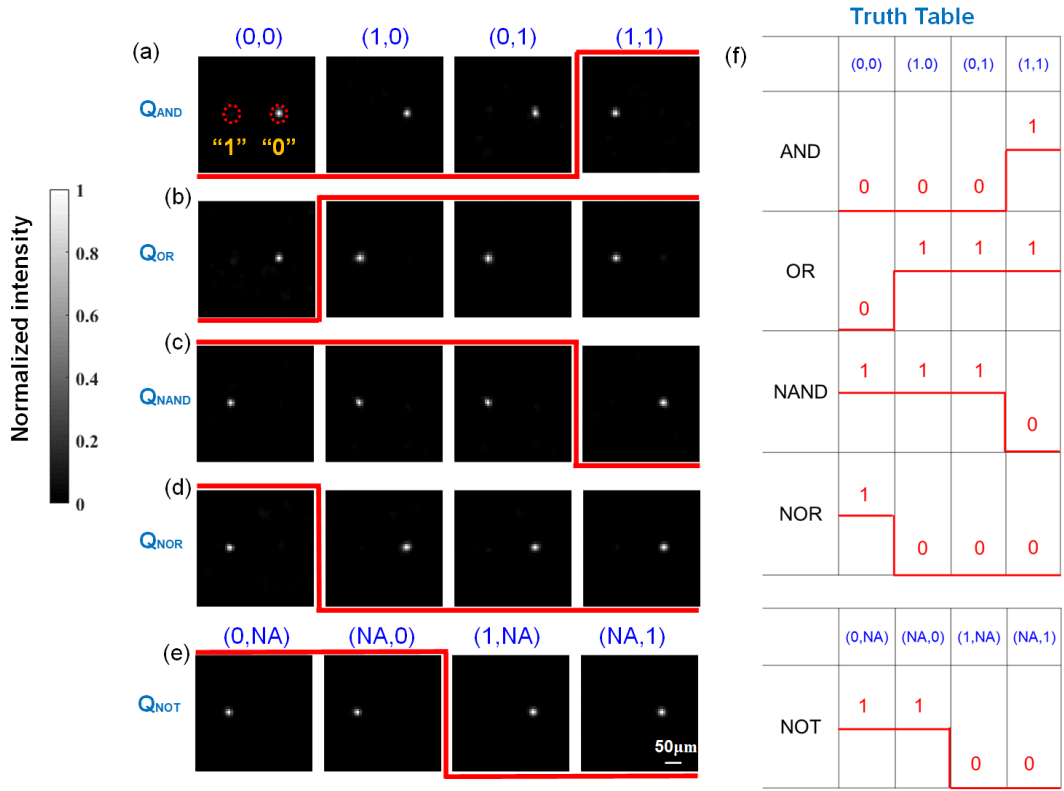

**Figure 5.** Experimental results of different logic operations. Logic gate digital output Boolean response at each of the four input states ((0, 0), (1, 0), (0, 1) and (1, 1) marked in blue) for (a) AND, (b) OR, (c) NAND, and (d) NOR logic gate; (e) Logic gate digital output Boolean response at each of the four states ((0, NA), (NA, 0), (1, NA) and (NA, 1)) for NOT logic gate; (f) The respective subregion inputs (blue) for each of the output states directly relate to the logic gates truth table of inputs. Digits "0" and "1" marked in yellow in the first Figure represent the logic states. Intensities are normalized to the maximum of each Figure. "(0, NA)" marked in blue indicates that only the subregion representing input binary digit "0" on the leftmost column is switched on. Other optical logic operations in this group ("NOT" logic gate) have similar procedures. $Q_{AND}$-$Q_{NOT}$: logic output states. The red lines below and above each Figure in Figure 5a-e represent the optical logic output states of "0" and "1", respectively.

## 4. Discussions

Thus far, we have described the concept and the working principle to achieve basic logic operations with diffused light through a scattering medium using transmission matrix-based wavefront shaping. Experimental results have demonstrated the feasibility and robustness of the proposed method. A few more things need to be discussed herein.

### 4.1 Speed optimization

Currently, the measurement of transmission matrix costs ~37 seconds, which is mainly restricted by the working frame rate (500 Hz) of the camera used in the system. One effective way to shorten the duration of TM measurement is to adopt a faster detector, such as a photoelectric detector[34], to achieve a measurement time of ~800 milliseconds. The significantly increased speed can also make it compatible for a real complex or dynamic environment such as fog. The computing speed is also determined by the speed of the DMD (23 kHz), when it modulates and delivers light from different subregions on the DMD through the scattering medium.

### 4.2 Selection of sensor regions to present the logic states

The selection of sensor regions on the output plane to present the logic states also matters. As known, the focal enhancement ratio pattern on the output plane is highly correlated with the intensity distribution profile of the speckle grains formed by the reference light: [35] at the dark grain spots, the focal enhancement ratio is low; at the bright grain spots, the focal enhancement ratio is high. This effect leads large variations among intensities of focuses formed in different areas on the output plane, which further affects the viability of the proposed logic operations. Therefore, cautiously selecting two bright grains as the sites of focuses can increase the robustness of the proposed logic operations. On other hand, several methods can be used to improve the success rate. For example, a super-pixel method can be adopted to achieve more uniform distribution of intensity of the resultant optical focuses by using the DMD as a phase modulation device.[39] Another method is to scan the incident laser beam so as to achieve maximum optical transmission through the scattering medium before the TM measurement.[35]

### 4.3 Realization of all seven basic logic gates

It should be noted that the proposed scheme can, in principle, directly construct all seven basic logic functions (including XOR and XAND that are not yet reported in this study). This can be done by changing the intensity of focus with a finer resolution; the focal intensity can be adjusted in a large range by blocking certain portion of the constructive channels, while the phase value maintains largely unchanged, as long as the remaining channels are sufficient to make the focus to be in phase or opposite phase with the common reference. This method can also be used to improve the intensity contrast ratio as discussed in Figure 5. Note that, however, it requires more accurate computation to determine the desired intensity and phase of each focus.

### 4.4 Scalability and reconfigurability

The current scheme can be extended for more functions, such as achieving multi-bit logic operations. It can be achieved by, for example, cascading multiple current logic gates in a design discussed in Supporting Information 6. In experiment, we only used nine sub-TMs with dimension of 1024×40000 to realize the five basic optical logic gates (AND, OR, NOT, NAND, and NOR) with all light from the DMD illuminating the ground glass. The ground glass, theoretically, can support many groups of such optical logic gates due to the high rank of its transmission matrix. Moreover, the extension will not considerably increase the complexity for TM measurement thanks to the inverse design approach. The other factor that needs to be considered is the pixel number of the DMD used in the study, which is 1080×1920. Thus, it can support up to 10 logic gates with the current experiment conditions. Or, if no pixels are binned to form mega pixels, this DMD can support up to 220 logic gates. That said, multiple DMDs in combination or a DMD with more pixels is desired to achieve the mentioned large-scale cascaded optical logic gates. The dimension of the DMD will not be a major bottleneck to the extension potential of our work, especially when we consider the rapid development of DMD or other spatial light modulators in the past two decades and in the near future.

The inverse approach is a common strategy for the design of existing optical logic components, but it will encounter some limitations because of fabrication inaccuracy and environmental disturbance; packaging the whole system in a black box to be less sensitive or immune to perturbations is usually required. When the scene of application varies, new design and/or fabrication of components as well as system packaging might be necessary. For our approach, when the system or the environment is altered, the issue can be solved by measuring the TM again, which currently costs seconds but can be considerably shortened to less than 1 second. Such a reconfiguration capability is unmatched by any of existing optical logic schemes.

## 5. Conclusion

To conclude, optical computing through a scattering medium based on transmission matrix-based wavefront shaping are proposed and demonstrated in this work. Light is first modulated by a DMD with precalculated patterns and then it transmits through a ground glass diffuser to output logical states. Choosing a common reference speckle field for light from nine DMD subregions, we have experimentally demonstrated the feasibility and robustness of five basic logic functions (AND, OR, NOT, NAND, NOR). It empowers reconfigurability to the design of optical logic gates for dynamic systems, which is unmatched by any of existing implementations in the field. Moreover, to the best of our knowledge, this is the first implementation of optical computing through scattering media. As the transmission matrix of the scattering medium/system can be measured instantly to adapt to environment perturbation or change, the method opens new venues towards reconfigurable optical computing in a complex or dynamic environment.

**Supporting Information**

Supporting Information is available from the Wiley Online Library or from the author.

**Acknowledgements**

Z.Y., Y.S and T.Z. contributed equally to this work. The work was supported by National Natural Science Foundation of China (NSFC) (81930048, 81627805, 81671726), Hong Kong Research Grant Council (15217721, R5029-19, 25204416), Hong Kong Innovation and Technology Commission (GHP/043/19SZ, GHP/044/19GD, ITS/022/18), Guangdong Science and Technology Commission (2019A1515011374, 2019BT02X105), and Shenzhen Science and Technology Innovation Commission (JCYJ20170818104421564).

**Conflict of Interest**

The authors declare no competing financial interests.

# Supplementary material of

# Reconfigurable optical computing through scattering media with wavefront shaping

**Zhipeng Yu[1,2] †, Yuchen Song[1,2] †, Tianting Zhong[1,2], Huanhao Li[1,2], Wei Zheng[3], and Puxiang Lai[1,2]***

[1]Department of Biomedical Engineering, Hong Kong Polytechnic University, Hong Kong SAR

[2]Shenzhen Research Institute, Hong Kong Polytechnic University, Shenzhen 518057, China

[3]Research Laboratory for Biomedical Optics and Molecular Imaging, Shenzhen Institutes of Advanced Technology, Chinese Academy of Sciences, Shenzhen 518055, China

† Contributed equally to this work

*Corresponding author: puxiang.lai@polyu.edu.hk

**Supplementary Note 1: Optical setup**

A continuous wave laser source of 488 nm wavelength (OBIS LS, Coherent, USA) serves as the light source, and the maximum output power is 150 mw. Light is first expanded by a 4f system (lens L1 from Thorlabs, AC254-30-A, f = 30 mm and lens L2 from Thorlabs, AC254-200-A, f = 200 mm) to illuminate the whole screen of the DMD (V-9501, Vialux GmbH, Germany). After being modulated and reflected by the DMD, light is shrunk by another 4f system (lens L3 from Thorlabs, AC254-200-A, f = 200 mm, and L4 from Thorlabs, AC254-30-A, f = 30 mm), and then it is converged by an objective lens (40×/0.65, Daheng Optics, China) onto a ground glass diffuser (DG10-120, Thorlabs, USA). At last, the output optical field is recorded by a CMOS camera (BFS-U3-04S2M-CS, FLIR Integrated Imaging Solutions Inc., Canada), which is triggered by the DMD. In experiment, once the DMD pattern is updated, the camera will record an optical pattern.

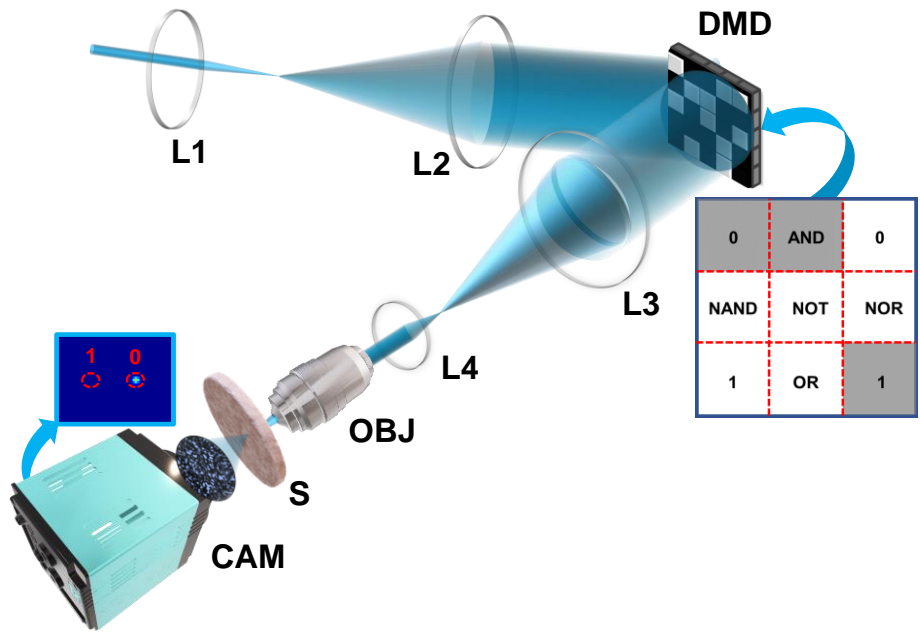

Fig. S1 Optical setup. CAM: CMOS camera; DMD: digital micromirror device; L1-L4: lens; OBJ: objective lens; S: ground glass diffuser as the scattering medium. The figure presents a logic output state of "0" for logic operation "0·1" (with DMD Subregions 0, AND, and 1

activated).

**Supplementary Note 2: Intensity and phase control for optical focuses through scattering media**

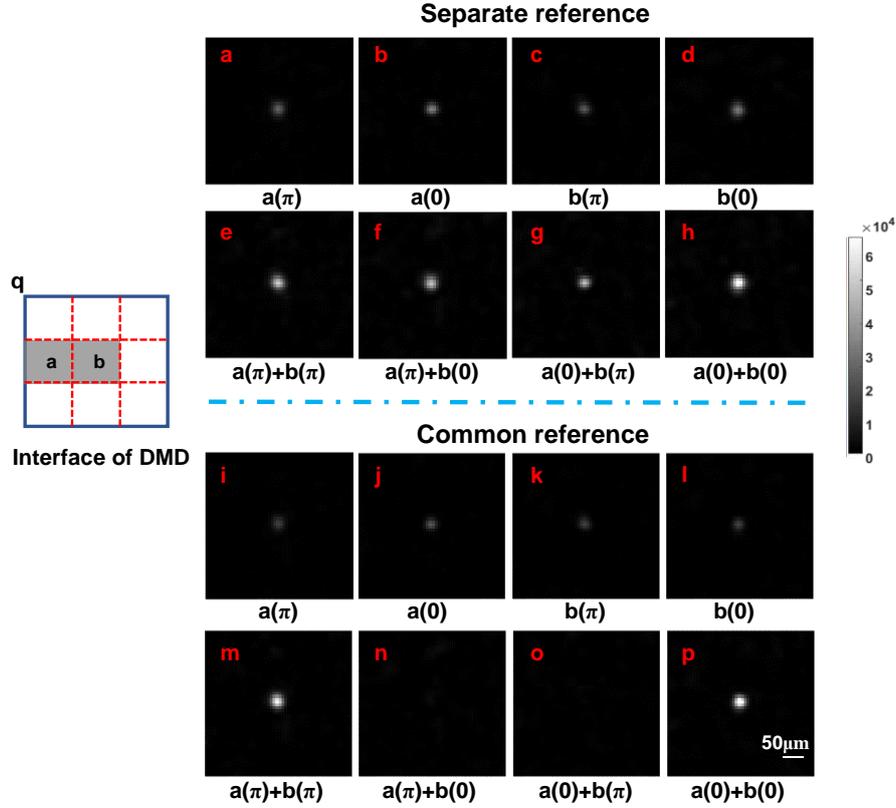

Fig. S2 Illustration of focuses generated from different subregions with separated or common reference as well as their interference. (a)-(d): Optical focus, which is in phase or opposite phase with the reference, is achieved with a separate reference for each subregion; (e)-(h): Results of interference between focuses generated from Subregions a and b shown in (a)-(d); (i)-(l): Optical focus, which is in phase or opposite phase with the reference, is achieved with a common reference for all subregions; (m)-(p): Results of interference between focuses generated from two subregions shown in (i)-(l); (q) The arrangement of subregions on the DMD screen. Two subregions marked in grey serve as the working regions, denoted as "a" and "b", respectively. "a(0)" indicates the focus formed by the subregion is in phase with the reference; "a(0)+b(π)" represents the inference between one focus (in phase with the reference) generated from Subregion "a" and one focus (in opposite phase with the reference) generated from Subregion "b".

Two subregions marked in grey on the DMD are used and renamed as Subregions a and b, respectively, as shown in Fig. S2q. Each sub-TM of the scattering medium corresponding to these two subregions will be calculated in sequence. The experimental verification can be found in Fig. S2. "0" in the bracket means the focus is in phase with the reference and "π" in the bracket means the focus is in opposite phase with the reference. Two groups of experiments were performed. The first group adopted separate reference for each subregion (with all pixels in the corresponding subregion switched "on"). The resultant optical focuses as well as their interference with subregions in phase or opposite phase to the reference are shown in Figs. S2a-h. The second group adopted the aforementioned common reference for all subregions, and the corresponding results are shown Figs. S2i-p. As seen, with separate references, constructive and destructive interference cannot be observed at the same time for different combinations of phase relationship between optical (focal) fields from two subregions. This indicates that there is no explicit phase difference (not in phase or in opposite phase) between the two focuses formed by different subregions. The results are chosen from 100 groups of experiment data. Statistics results suggest a state of complete disorder in different areas on the output plane, shown in Fig. S3. By contrast, with a common reference, there is constructive interference, as represented by enhanced focal intensity, when the two focuses are both in phase or both in opposite phase with the common reference, and there is destructive interference when one focus is in phase but the other is in opposite phase with the common reference. This shows the importance and necessity to regulate the phase of focuses formed by various subregions with respect to the common reference. Similarly, the results are chosen from 100 groups of experiment data; for statistical analysis, please refer to Fig. S3.

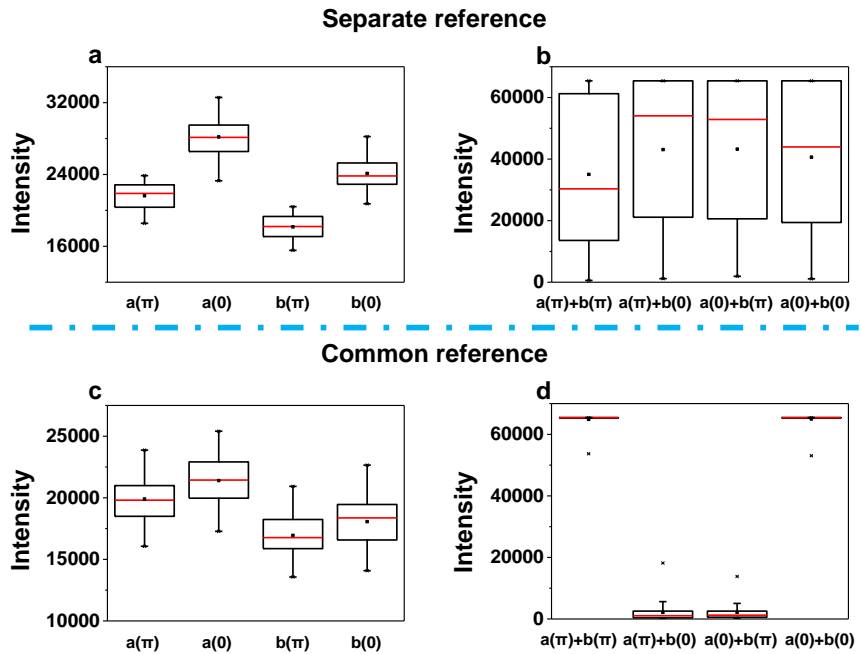

Fig. S3. Statistics of intensities of focuses in Figs. S2(a-p) in the main text. Results with separate reference are shown in (a) and (b), corresponding to before and after field interference of two subregions, respectively. The intensities of focuses after interference distribute in a large range, indicating the lack of an explicit phase difference among focuses generated by two subregions. Results with a common reference are shown in (c) and (d), corresponding to before and after field interference of two subregions, respectively. The sharp contrast of focal intensities (interference between fields of same phase versus interference between fields of opposite phase) confirm that the desired phases of focuses generated by different subregions are accurate and hence suitable for constructive and destructive interference.

**Supplementary Note 3: Interference of dual-focus optical fields formed by individual DMD subregions**

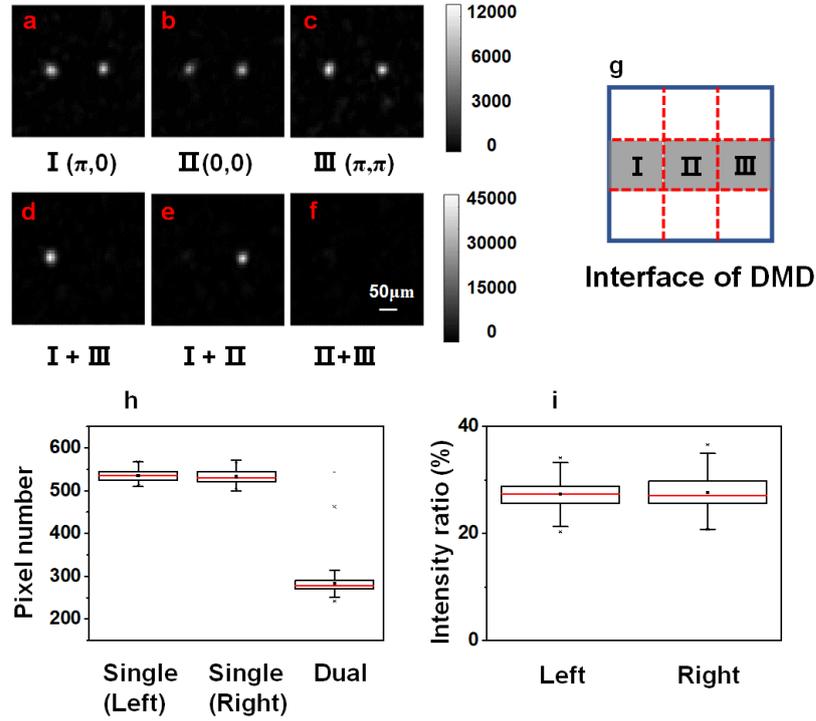

Fig. S4 Illustration of the phase and intensity control before and after interference of optical fields formed by individual DMD subregions. (a) Optical focus on the left of the output plane is in opposite phase and focus on the right is in phase with the common reference; generated by Subregion I. (b) Both optical focuses are in phase with the reference; generated by Subregion II. (c) Both optical focuses are in opposite phase with the reference; generated by Subregion III. (d-f) Interference of optical fields generated by any of two Subregions. For example, "I+II" represents the interference on the camera plane between optical fields by Subregions I and II. Note that (a)-(c) share the same color bar, (d)-(f) share the same color bar, and (a)-(f) share the same scale bar. (g) The arrangement of the demonstration three subregions on the DMD. (h) Number of pixels in Subregion II that should be switched "on" in order to generate a single focus on the left, a single focus on the right and dual focuses, respectively, after the intersection operation. (i) Ratios of intensity of the two focuses with respect to that of the focus before the intersection operation. Red lines indicate the median values.

As demonstrated in the manuscript, channels that interfere destructively with the reference field should be blocked to highlight the constructive interferences and hence to make the focus in phase with the reference. In our experiment, even though the intersection operation will weaken the intensity of each of the two focuses, there are still a large portion of channels that interfere constructively with each other, forming focuses that are in phase or opposite phase with the reference. Also, it should be noted that there might be crosstalk between the resultant two focuses as they originate from the same DMD subregion. But how this effect influences the phase and intensity of the two focuses needs further exploration.

The experimental verification is given in Fig. S4. Three subregions on the DMD screen, marked in grey in Fig. S4g, are used and referred as Subregions I, II, and III (distinctive from Subregions a and b in Fig. S2q). They are selected from the nine subregions; each subregion can generate two optical focuses in the designated areas on the output plane with desired phases, as exampled in Figs. S4a-c. The desired phases of the focuses formed by Subregion I are π (left) and 0 (right), those by Subregion II are 0 (left) and 0 (right), and those by Subregion III are π (left) and π (right). Interference of different combinations of these three optical fields (*i.e.,* I+III, I+II, and II+III) are shown in Fig. S4d-f. As seen, in Fig. S4d, a brighter focus is observed on the left but the focus on the right is suppressed, indicating that there is constructive interference (π and π) for the left focus but destructive interference (0 and π) for the right focus. In Figs. S4e and 4f, consistent results are obtained. These results confirm that desired phases are correctly designated, and the intersection operation does not affect the phase of the optical focuses. Statistics about the intensity of different focuses in Figs. S4a-f can be referred to Fig. S5.

Next, let us examine how the intersection operation affects the intensity of optical focuses. We selected Subregion II as the working unit, and randomly picked 100 pairs of areas on the output plane. As seen, almost half of the pixels in Subregion II are switched "off" (active pixels reduced from around 520 down to around 260) after the intersection operation, as shown in Fig. S4h. As the focal intensity is linearly correlated with the effective pixel number [1], the resultant focal intensity is expected to decrease by half. As the remaining pixels are used to contribute to two focuses, intuitively intensity of each focus is decreased by another half. Note that, however, there might be competition of energy between the two focuses during the whole transmitting process as photons are from the same DMD subregion. To check that, we obtained the statistics of experimental focal intensities before and after the intersection operation. As shown in Fig. S4i, no significant imbalance of focal intensity is found between the two focuses, and the intensity of each of the two focuses is approximately one quarter of that of the focus before the intersection operation (amplitude of the electric field of the focus almost halves), agreeing quite well with the deductions above.

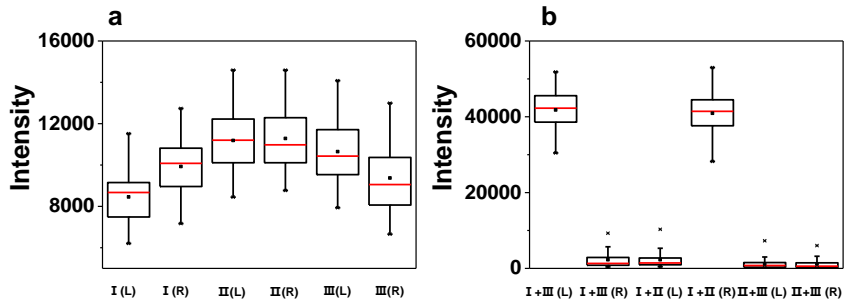

Fig. S5. Statistics of intensities of the dual focuses shown in Figs. S4(a-f) in the main text.

Fig. S5a shows the intensity distribution of the dual focuses from the three DMD subregions as shown in Fig. S4g in the main text after the intersection operation. As seen, no significant imbalance of focal intensity is found among all these focuses; the moderate difference may attribute to the variations of enhancement efficiency for different DMD subregions and different output plane positions. Fig. S5b shows the intensity distribution of interfered fields between any two of the dual-focus patterns as shown in Fig. S4 (a-c). As seen, when the desired phases of focuses are the same, such as I+III (L) and I+II (R), the focus intensity can be increased by ~4 times, confirming the left focuses formed by I and III are accurately in phase (both have phase of π), and so are the right focuses formed by II and II (both have phase of 0). On the contrast, when the desired phases of focuses are opposite, the focal intensities after interference are almost zero, confirming these focuses are in opposite phase (*i.e.,* one is π and the other is 0). These results confirm that desired phases are correctly designated, and the intersection operation does not affect the phase of the optical focuses.

**Supplementary Note 4: Phase and intensity calculation of focuses**

| Focus area \ Active unit | 0 (left) | 1 (left) | 0 (right) | 1 (right) | AND | OR | NOT | NAND | NOR |
|---|---|---|---|---|---|---|---|---|---|
| "0" | a | b | c | d | e | f | g | h | i |
| "1" | j | k | l | m | n | o | p | q | r |

Table S1: Desired phases and intensities of focuses from each DMD subregion in an inverse design. Characters "a" to "r" are the required electric field amplitudes of focuses at the focus areas representing logic states "0" and "1". If the character is negative, the focus is in opposite phase with the reference; otherwise, the focus is in phase with the reference.

Characters "a" to "r" represent the electric field amplitude of focuses generated by the nine DMD subregions/control units. The combination of different groups can act as different logic operations. And their relationship should obey the below inequations:

$$
\begin{aligned}
&|a+c+e| \geq 2*|j+l+n| \quad |a+c+f| \geq 2*|j+l+o| \\
&|a+d+e| \geq 2*|j+m+n| \quad 2*|a+d+f| \leq |j+m+o| \\
&|b+c+e| \geq 2*|k+l+n| \quad 2*|b+c+f| \leq |k+l+o| \\
&2*|b+d+e| \leq |k+m+n| \quad 2*|b+d+f| \leq |k+m+o|
\end{aligned}
$$

$$
\begin{aligned}
&2*|a+c+h| \leq |j+l+q| \quad 2*|a+c+i| \leq |j+l+r| \quad 2*|a+g| \leq |j+p| \\
&2*|a+d+h| \leq |j+m+q| \quad |a+d+i| \geq 2*|j+m+r| \quad |b+g| \geq 2*|k+p| \\
&2*|b+c+h| \leq |k+l+q| \quad |b+c+i| \geq 2*|k+l+r| \quad 2*|c+g| \leq |l+p| \\
&|b+d+h| \geq 2*|k+m+q| \quad |b+d+i| \geq 2*|k+m+r| \quad |d+g| \geq 2*|m+p|
\end{aligned}
$$
(S1)

If a character is negative, the phase of the corresponding focus is π (in opposite phase with the reference); if a character is positive, the phase of the corresponding focus is 0 (in phase with the reference). The threshold setting is if logic output is "0", the absolute amplitude value of the synthetic field in the area of logical state "0" should be twice larger than that in the area of logical state "1", and vice versa.

**Supplementary Note 5: Statistics of intensity contrast of logic operations with 200 pairs of areas on the output plane to present the logic states**

In order to further demonstrate the reliability of the method, 200 pairs of areas on the output plane were chosen in experiment as the focuses formed by light from individual DMD subregions and to represent the logic "1" and "0" outputs (it contains 200 groups of data and each group contains 20 different logic operations). As shown in Fig. S6a, the number of logical operations with intensity contrast values below 3 dB are 216 (including the invalid sets), which is about 5.4% of the total operations (4,000). From Fig. S4b, the average intensity contrast values of all types of logical operations are over 5 dB except $0\oplus0$, which has an average value slightly below 5 dB.

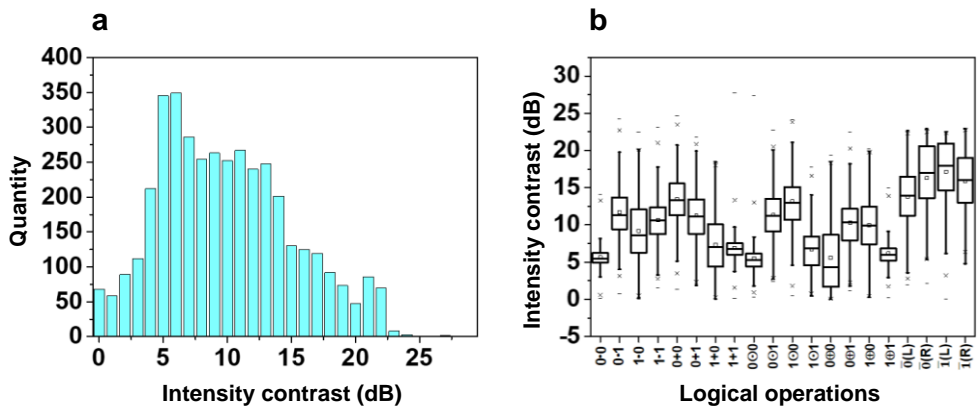

Fig. S6 200 pairs of areas on the output plane were chosen in experiment as the focuses formed by light from individual DMD subregions and to represent the logic "1" and "0" outputs. (a) Statistical distribution of intensity contrast of all logical operations (the total number is 200×20). The logic operation is invalid when the intensity contrast is zero. Intensity contrasts are rounded down along the horizontal axis. (b) Distribution of intensity contrasts of certain type of logical operation from the 200 groups of data.

**Supplementary Note 6: Cascaded optical logic gates for function extension**

The DMD screen can be divided into a large number of subregions or control units to represent various logic gates and connection buttons. One logic gate is the basic block to achieve a group of basic logic operations as demonstrated in this study. Connection buttons contain three logic types "AND", "OR" and "NOT" to connect adjacent logic gates to achieve cascaded function. For example, for logic operation "$(0\oplus0)+(1+0)$", there will be two logic gates, with one functioning as "$0\oplus0$" and the other as "$1+0$", and one connection button to provide function "OR". In operation, the corresponding control units on the DMD will display the pre-calculated wavefronts obtained with our TM-based wavefront shaping method using an inverse design. A typical DMD has 1080×1920 pixels, allowing for extension towards complicated (including more functions and more bits) logic operations.

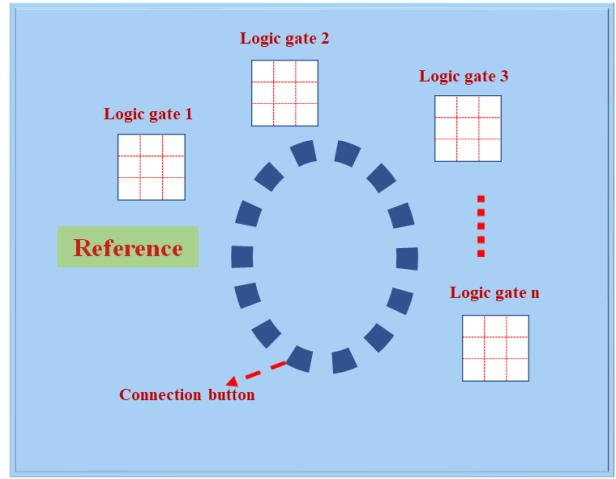

Fig. S7 Arrangement of the DMD for cascaded optical logic gates

1. I. M. Vellekoop and A. Mosk, "Focusing coherent light through opaque strongly scattering media," Opt. Lett. **32**, 2309-2311 (2007).